\documentclass[aps,pra, twocolumn,nofootinbib, showpacs]{revtex4-1} 
\usepackage[unicode=true,pdfusetitle, bookmarks=true,bookmarksnumbered=false,bookmarksopen=false, breaklinks=false,pdfborder={0 0 0},backref=false,colorlinks=false] {hyperref}
\hypersetup{ colorlinks,linkcolor=myurlcolor,citecolor=myurlcolor,urlcolor=myurlcolor}
\usepackage{braket,colortbl,amsthm,amsmath,cleveref,amssymb,txfonts}\definecolor{myurlcolor}{rgb}{0,0,0.7}
\usepackage{graphics,graphicx}
\usepackage{color}
\usepackage{amssymb}
\usepackage{amsthm}
\usepackage{amsfonts}
\usepackage{float}
\usepackage{graphicx}
\usepackage[format = plain,labelfont = bf,up, textfont = normal , up, justification =raggedright, singlelinecheck =false]{caption}
\usepackage{subcaption}
\usepackage{tabularx}
\usepackage{amsmath}
\usepackage{braket}

\usepackage{graphicx}
\usepackage[utf8x]{inputenc}
\usepackage{color,soul}
\usepackage{amsmath}
\usepackage{braket}
\usepackage{latexsym}
\usepackage{amssymb}
\usepackage{amsthm}
\usepackage{bm}
\usepackage{graphics,epstopdf}
\usepackage{color}\usepackage{amsmath}
\usepackage{enumitem}
\usepackage{fmtcount}
\usepackage{booktabs}
\usepackage{csquotes}
\usepackage{epsfig}
\usepackage{romannum}
\theoremstyle{plain}

\def\bea{\begin{eqnarray}}
\def\eea{\end{eqnarray}}
\def\ba{\begin{array}}
\def\ea{\end{array}}

\def\ket{\rangle}
\def\bra{\langle}
\def\beq{\begin{equation}}
\def\eeq{\end{equation}}

\begin{document}
\title{Beating detection loophole in nonlinear entanglement witnesses}

\author{Kornikar Sen, Sreetama Das, Ujjwal Sen}

\affiliation{Harish-Chandra Research Institute, HBNI, Chhatnag Road, Jhunsi, Allahabad 211 019, India}
\begin{abstract}
Detectors in the laboratory are often unlike their ideal theoretical cousins. They have nonideal efficiencies, which may then lead to nontrivial implications. We show how it is possible to predict correct answers about whether a shared quantum state is entangled in spite of finite detector efficiencies, when the tool for entanglement detection is a nonlinear entanglement witness. We first consider the detection loophole for shared quantum states with nonpositive partial transpose. We subsequently find nonlinear witness operators for bound entangled states with positive partial transpose, and show how the detection loophole can be closed also in such instances.
\end{abstract}
\maketitle

\section{Introduction}

Entanglement is a useful resource in quantum tasks 
\cite{HHHH,sree}, 
including quantum teleportation \cite{tele},
quantum
dense coding \cite{dc}, 
and entanglement-based quantum cryptography
\cite{ekert}. 
 It is therefore important to find out whether a shared
quantum state is entangled. There are several methods known for detection
of entanglement, including the positive partial transpose (PPT) criterion
\cite{Peres,Horodecki},
entropic criteria \cite{entr1,entr2},
Bell inequalities \cite{Bell}, 
and entanglement witnesses
\cite{Horodecki,witness}. 
A necessary and sufficient criterion that
is analytically tractable or numerically efficient remains elusive. There
have been significant advances in experimental detection of entanglement
by using the above criteria \cite{exp,wit_exp}.

Whatever is the approach for detecting entanglement, it will of course
involve measurements on the shared quantum state. The devices that are
used for such measurements are typically assumed, in theoretical
discussions, to be ideal.

From an experimental perspective, a useful method for detecting
entanglement is by using entanglement witnesses, which are linear
operators on the space of quantum states (density matrices) and which
provide a sufficient condition for detecting entanglement. The criterion
is based on the Hahn-Banach separation theorem on normed linear spaces
\cite{hahn}. A large number of experiments have
utilized entanglement witnesses for detecting entanglement
\cite{wit_exp}.

Bell inequality violation for a shared quantum state implies that the
state cannot be described by a local hidden variable model. It also 
implies that the state is entangled. Indeed, a typical Bell inequality, e.g., the Clauser-Horne-Shimony-Holt inequality \cite{CHSH},
is a nonoptimal entanglement witness.
There exists a series of works on the detection loophole for Bell
inequality violations \cite{X} (see also \cite{two}), where the theoretical discussion allows the detectors to have nonideal
efficiencies. Experimental violation of Bell inequalities, while acknowledging nonideal detector efficiencies, has been explored in several works
\cite{exp_bell}. Reference
\cite{Bruss} 
considered implications
of the detection loophole in experiments for entanglement detection via
entanglement witnesses.

Entanglement witnesses predicted by the Hahn-Banach theorem are linear
operators. For every entangled state, there always exists an entanglement
witness that can detect it, as well as some -- but not all -- other
entangled states. However, it is possible to add nonlinear terms to linear
witness operators that detect the entangled states that are detected by
the linear parent witness, as well as some more entangled states
\cite{nonlinear1,nonlinear2,nonlinear3}.

There are two results obtained in this paper, and in the first one we
find limits on the threshold efficiency of detectors for implementing
nonlinear entanglement witnesses, for entangled states with a nonpositive partial transpose (NPPT).

The second one relates to bound entangled states, which, in the two-party
case, are shared quantum states that are entangled but not distillable,
i.e., it is not possible to obtain singlets, even asymptotically, from the
shared state by local quantum operations and classical communication \cite{bound}.
In this part, we begin by constructing
nonlinear entanglement witnesses for bound entangled states with positive
partial transpose. As a particular example, we consider nonlinear
witnesses for the family of bound entangled states given in Ref.
\cite{eita}. We subsequently provide
bounds on the threshold efficiency of detectors for detecting the bound
entangled state by utilizing the nonlinear witness.

The paper is arranged as follows. In Sec. \ref{sec2}, we briefly discuss
certain general aspects of linear and nonlinear entanglement witnesses.
The detection loophole for linear witnesses is reviewed in Sec. \ref{sec3},
which also sets up the notations for the succeeding sections. We present
our results on the detection loophole for nonlinear entanglement witnesses
for entangled states with a NPPT in Sec.
\ref{sec4}. Bound entangled states with PPT are
considered in Sec. \ref{sec-5}, where we first present nonlinear witnesses for
them, and then consider the limits on detection efficiencies for their
detection using nonlinear witnesses. We present a conclusion in Sec.
\ref{sec6}.

\section{LINEAR AND NONLINEAR ENTANGLEMENT WITNESSES}
\label{sec2}

Among the various methods for detecting entangled states, there are a few which can be realized experimentally without going through an entire state tomography. One of them is by using witness operators.  The concept of the entanglement witness is based on the Hahn$-$Banach theorem \cite{hahn}. It states that if $S$ is a closed and convex set in a normed linear space $L$, and $x\in L \setminus S$, then there exists a continuous functional $f:L\rightarrow\mathbb{R}$ such that $f(s)<r\leq f(x)$ for all $s\in S$ where $r\in \mathbb{R}$. The space of density matrices on a given Hilbert space forms a normed linear space for the norm, $||\rho||=\sqrt{\text{tr}(\rho \rho^\dagger)}$, of a density matrix $\rho$. This remains valid for density matrices on the tensor products of several Hilbert spaces, and in particular for the tensor product, of two Hilbert spaces $H_A$ and $H_B$. We now identify $S$ with the set of separable states on $H_A\otimes H_B$, and $x$ with an entangled state \cite{HHHH,sree} on the same bipartite system. We note that separable states form a closed and convex set in the space of density matrices. The Hahn-Banach theorem, therefore, guarantees the existence of a functional which separates the set of separable states with the entangled state. This functional is called a witness operator \cite{witness} and is defined as an operator $W$ which satisfies the following conditions:
\begin{eqnarray}
 \mbox{tr}(W\rho_s)&\geq &0\text{ for all } \rho_s\in s, \label{eq5} \nonumber \\ 
  \mbox{tr}(W\rho)&<&0\text{ for at least one entangled state }\rho. \nonumber \label{eq6}
\end{eqnarray} 
Note that if for any state $\rho$ one gets tr$(W\rho)<0$ one can surely conclude that it is entangled. Moreover, since the set of nonseparable states is open, there will always exist an open ball, in a suitable metric, the entanglement of every state of which will be detected by the same witness. This is a useful fact for experimental implementation of the witness, as small and often inevitable errors in the preparation of the state can then be nullified. Furthermore, for every entangled state, $\rho$, there always exists a witness that detects it. An example of a witness operator for an NPPT state $\rho_\phi$ is $ W_{\phi}=|{\phi}\ket\bra{\phi}|^{T_B}$ \cite{witness}, where $|{\phi}\ket$ is an eigenvector corresponding to a negative eigenvalue of $\rho_\phi^{T_B}$. Here, one can easily check that the expectation value of $W_{\phi}$ is positive for all separable states and negative for $\rho_\phi$, i.e., it can detect the entanglement of $\rho_\phi$. But such witness operators can only detect NPPT states. Witness operators for detecting PPT bound entangled states are discussed in  Sec. \ref{sec-5}.  
\par
 The operator $W$ is a \enquote{linear} operator, in the sense that it acts linearly on the space of density matrices. One can get more efficient witness operators by adding nonlinear terms to linear witness operators in such a way that the new \enquote{nonlinear witness operator} can detect the entangled states that can be detected by the parent linear witness operator, as well as additional ones. We will introduce nonlinear witness operators more formally in Sec. \ref{sec4}.

\section{DETECTION LOOPHOLE}
\label{sec3}
In this section, we briefly recapitulate the implications of a finite (i.e., nonzero) efficiency for linear entanglement witnesses \cite{Bruss}. While we consider only the two-qubit case, the methods work also in higher dimensions and higher number of parties. A decomposition of the witness operator, $W$, in the two-qubit case, is given by
\begin{eqnarray}
W&=&C_{00}I\otimes I+\sum_{i=1}^3C_{0i}I\otimes \sigma_i \nonumber \\
&&+\sum_{i=1}^3C_{i0}\sigma_i\otimes I+\sum_{i,j=1}^3C_{ij}\sigma_i\otimes\sigma_j \nonumber \\
&=&C_{00}I\otimes I+\sum_{k=1}^{15} C_kS_k. \label{eq28}
\end{eqnarray} 
where $S_k$'s are tensor products of all combinations of two $\sigma_i$ $(i=0,1,2,3)$ except $I\otimes I$, with $\sigma_0=I$, $\sigma_i$ for $i=1,2,3$ being the Pauli matrices. Here, $I$ is the identity operator on the qubit Hilbert space. $C_{ij}$ and $C_k$ are real numbers. To detect the entanglement of a two-qubit state through the expectation value of $W$ in that state, one has to measure these $S_k$'s for that state. Since there could be errors in these measurements, the status of a state - with respect to whether or not it is entangled - found by using the value of a witness operator could have a \enquote{loophole} in the argument. We want to find the condition for overcoming such a loophole. The measured expectation value of $S_k$ for a certain two-qubit state, $\rho$, is given by $\langle S_k \rangle _m =\frac{\sum n_i \lambda_i}{N}=\frac{ \sum ({\tilde{n}}_i+\epsilon_{+i}-\epsilon_{-i})\lambda_i}{\tilde{N}+\epsilon_+ -\epsilon_-}$. 
 Here, $n_i$ denotes the number of times that the $i$th eigenvalue $\lambda_i$ of $S_k$ has clicked in experiment, and $\tilde{n}_i$ denotes the number of times the same should have clicked in case of perfect detectors. Also, $N=\sum_i n_i$ and $\tilde{N}=\sum_i \tilde{n}_i$.  $\epsilon_{+i}$ are the number of additional events and $\epsilon_{-i}$ are the number of lost events at the $i$th outlet. The total number of additional and lost events are, respectively, given by $\epsilon_+=\sum_i \epsilon_{+i}$ and $\epsilon_-=\sum_i\epsilon_{-i}$, and the corresponding detection efficiencies are defined as $\eta_+=\frac{\tilde{N}}{\tilde{N}+\epsilon_+}$ (equal to additional event efficiency) and $\eta_-=\frac{\tilde{N}-\epsilon_-}{\tilde{N}}$ (equal to lost event efficiency).
    In this paper, we assume that the additional event efficiency $\eta_+=1$, i.e., $\epsilon_+=0$, and that the $\epsilon_{-i}$'s are equal for all $i$'s and the value is, say, $\epsilon$. With additional notations and algebra, those assumptions can of course be lifted. With these assumptions, we have $\langle S_k \rangle _m =\frac{\sum\tilde{n}_i-\epsilon\sum\lambda_i}{\tilde{N}-\epsilon_-}$.
    Since $S_k$'s are tensor products of the Pauli matrices (which are all traceless matrices) with each other or with the identity matrix, the traces of $S_k$'s are zero. Hence, we get $\langle S_k\rangle _m=\frac{1}{\eta_-}\frac{\sum \tilde{n}_i\lambda_i}{\tilde{N}}=\frac{1}{\eta_-}\langle S \rangle _t$. 
Here, $\langle S_k \rangle _t$ denotes the true value of $S_k$, i.e, the expectation value of $S_k$ when measured with ideal detectors, for the state $\rho$. Now, from equation \eqref{eq28}, we have $\langle W \rangle _m=C_{00}+\frac{1}{\eta_-}\sum_k \langle S_k \rangle _m=C_{00}\left(1-\frac{1}{\eta_-}\right)+\frac{1}{\eta_-}\langle W \rangle _t$. An entangled state would be detected when $\langle W \rangle_t <0 $, so that we need
    \begin{equation}
    \langle W \rangle _m < C_{00}\left(1-\frac{1}{\eta_-}\right). \label{eq9} \nonumber 
    \end{equation}
  Now, for a particular detector, the value of $\eta_-$ is known, or can be estimated, usually, by independent means. If the measured value of the witness satisfies the above inequality for some state $\rho$, then, in spite of the inefficiencies of the detectors, we can conclude that the state $\rho$ is entangled. We can see from the relation that if one uses a witness such that in its decomposition the coefficient $C_{00}=0$ then the loophole in the detection cannot affect the result.

     Let us now take a particular witness operator, given by $W_{\phi^+}= |{\phi ^+}\ket \bra{\phi^+}| ^{T_B}$ (where $|{\phi^+}\ket=\frac{1}{\sqrt{2}}(|{00}\ket+|{11}\ket)$. In an ideal scenario, this witness operator will be able to detect the entanglement in any two-qubit state $\rho_{\phi^+}$ that has $|{\phi^+}\ket$ as the eigenvector corresponding to the negative eigenvalue of $\rho_{\phi^+}^{T_B}$. An exemplary family of such states is the Werner family \cite{1989}, $\rho_p =p|\psi^-\ket \bra \psi^-|+(1-p)\frac{1}{2}I\otimes \frac{1}{2}I$, for $\frac{1}{3}<p\leq 1$. See \cite{Peres, Horodecki} in this regard. Here, $|\psi^-\ket=\frac{1}{\sqrt{2}}(|01\ket-|10\ket)$. Note here that two-qubit states can have at most a single negative eigenvalue after being partially transposed \cite{Ppt}. If we repeat the above calculation with this witness operator, we will get the following condition:
    \begin{equation}
    \langle W_{\phi^+} \rangle _m < \frac{1}{4}\left(1-\frac{1}{\eta_-}\right). \nonumber
    \end{equation}
    Now, for example, if the lost event efficiency $\eta_->\frac{1}{3}$, then to overcome the loophole and detect an entangled state, one needs $\langle W_{\phi^+} \rangle _m<-\frac{1}{2}$.
 \section{DETECTION LOOPHOLE IN NONLINEAR WITNESS OPERATORS}
\label{sec4}
\begin{figure*}[t]
\includegraphics[scale=1.4]{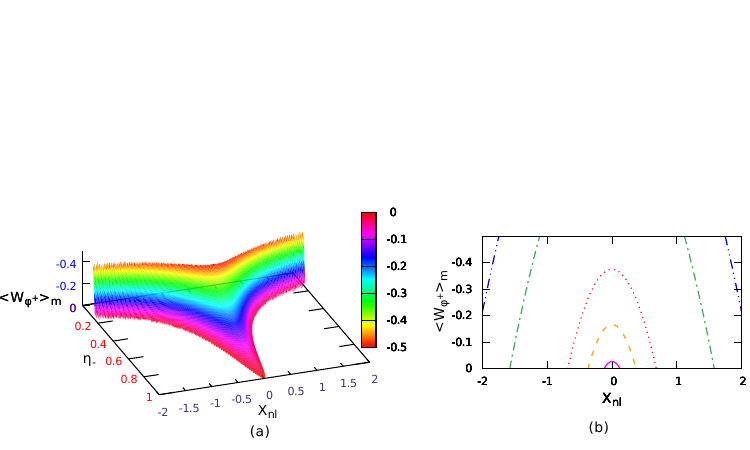}
\caption{Closing the detection loophole for a nonlinear entanglement witness. The values of the triad, $\langle W_{\phi^+}\rangle_m$, $X_{nl}$, and $\eta_-$, ascertain whether the entanglement in the state $\rho_{\phi^+}$ is detected, where $\rho_{\phi^+}$ is any state the partial transpose of which has $|\phi^+\ket$ as the eigenvector for its negative eigenvalue. In spite of the nonideal detector efficiency, the entanglement in $\rho_{\phi^+}$ is detected whenever the triad lies in the region above the surface plotted in panel (a). Here the different colors denote different ranges of values of $\langle W_{\phi^+}\rangle_m$ as indicated in the colorbox. In panel (b), values of $X_{nl}$ and $\langle W_{\phi^+}\rangle_m$ are shown for some fixed values of $\eta_-$. The blue double-dot-dashed, green dot-dashed, red dotted, orange dashed, and magenta continuous lines are, respectively, for $\eta_- = 0.15, 0.2, 0.4, 0.6,$ and 0.9. Each curve, therefore, is a cross section of the surface in panel (a) for different values of \(\eta_-\). The region outside each \(\eta_-\)-curve shows the values of $X_{nl}$ and $\langle W_{\phi^+}\rangle_m$ for which entanglement can be detected for that value of efficiency. It can be seen that as the efficiency increases, there is an increase in the  region for which successful detection of entanglement is possible. Note that $X_{nl}=\left( \langle H\rangle_m^2 +\langle A\rangle^2_m \right)^\frac{1}{2}$. All quantities are dimensionless.} 
\label{gr2}
\end{figure*}
As we have mentioned before, one can improve a linear witness operator by adding nonlinear terms to the linear witness operator, such that it \enquote{bends towards negativity}. If we consider the witness operator which witnesses NPPT states and is given by $|{\phi}\ket\bra{\phi}|^{T_B}$ then one can add a nonlinear term in the following way \cite{nonlinear1}:
\begin{equation}
F=\left<|{\phi}\ket\bra{\phi}|^{T_B}\right>-\frac{1}{s(\psi)}\left< X^{T_B}\right>\left<\left(X^{T_B}\right)^\dagger\right>, \nonumber
\end{equation}
where the expectations are for the state $\rho$, the entanglement of which we wish to detect.
Here, $X$ is given by $|{\phi}\ket\bra{\psi}|$, where $|{\psi}\ket$ is an arbitrary but fixed state and $s(\psi)$ is the square of the largest Schmidt decomposition coefficient of $|{\psi}\ket$. It is shown in \cite{nonlinear1} that $F\geq 0$ if the expectations in $F$ are for a separable state, and that when the expectations are for an entangled state $F<0$. Moreover, $F<0$ is true for more entangled states than for which $\langle W_\phi \rangle<0$. Here we wish to find the limits on the measured values of $F$ such that we can still correctly predict whether $\rho$ is entangled, in the case when the detectors are nonideal. To do this, one needs to find $F$, and hence has to measure $W$ and $X^{T_B}$, while acknowledging that the detectors are not ideal. Although $X^{T_B}$ is not Hermitian, we can decompose $X^{T_B}$ into Hermitian and anti-Hermitian parts as $X^{T_B}=H+iA$, where $H$ and $A$ are Hermitian, so that we get $\left< X^{T_B}\right>\left<\left(X^{T_B}\right)^\dagger\right>=\langle H \rangle^2+\langle A
 \rangle^2$. Since $H$ and $A$ are Hermitian, we can measure them. Here we have considered the case where all the operators are measured by using similarly engineered detectors so that the $\eta_-$ are the same for all the measurements. Just like $W$, the $H$ and $A$ can also be decomposed in terms of tensor products of the Pauli matrices and the identity matrix, and we obtain
 \begin{eqnarray}
 \langle H\rangle_m=C_{0H}\left(1-\frac{1}{\eta_-}\right)+\frac{1}{\eta_-}\langle H\rangle_t, \label{eq30} \\
 \langle A\rangle_m=C_{0A}\left(1-\frac{1}{\eta_-}\right)+\frac{1}{\eta_-}\langle A\rangle_t  \label{eq31}.
 \end{eqnarray}
 The suffixes $m$ and $t$ indicate, respectively, the measured and true values, and $C_{OH}=\frac{1}{4}$tr$(H)$ and $C_{OA}=\frac{1}{4}$tr$(A)$.
Hence, the measured value of the nonlinear witness operator is
\begin{eqnarray}
\langle F\rangle_m&=&\langle W_\phi\rangle_m -\frac{1}{s(\psi)}\left[\langle H\rangle_m^2+\langle A\rangle_m^2\right] \nonumber \\
&=&C_{00}\left(1-\frac{1}{\eta_-}\right)+\frac{1}{\eta_-}\langle W \rangle_t \nonumber \\
\ \ \ \ &&-\frac{1}{s(\psi)}\left[\langle H\rangle_m^2+\langle A\rangle_m^2\right] \nonumber \\
&=&C_{00}\left(1-\frac{1}{\eta_-}\right) \nonumber \\
&&+\frac{1}{\eta_-}\left[\langle F\rangle_t+\frac{1}{s(\psi)}\left(\langle H\rangle_t^2+\langle A\rangle_t^2\right)\right] \nonumber \\
 && -\frac{1}{s(\psi)}\left[\langle H\rangle_m^2+\langle A\rangle_m^2\right]. \nonumber
\end{eqnarray}
Putting the value of $\langle H \rangle_t$ and $\langle A \rangle_t$ in terms of $\langle H \rangle_m$ and $\langle A \rangle_m$ from \eqref{eq30} and \eqref{eq31}, we get
\begin{eqnarray}
\langle F\rangle_m&=&C_{00}\left(1-\frac{1}{\eta_-}\right)+\frac{1}{\eta_-}\langle F\rangle_t \nonumber \\
&+&\frac{\eta_-}{s(\psi)}\left( \langle H \rangle_m^2+k^2_H-2\langle H\rangle_mk_H \right) \nonumber \\
&+&\frac{\eta_-}{s(\psi)}\left(\langle A\rangle_m^2+k^2_A-2\langle A \rangle_mk_A\right)\nonumber \\
&-&\frac{1}{s(\psi)}\left[\langle H\rangle_m^2+\langle A\rangle_m^2\right], \nonumber
\end{eqnarray}
where $k_H=C_{0H}\left(1-\frac{1}{\eta_-}\right)$ and $k_A=C_{0A}\left(1-\frac{1}{\eta_-}\right)$.
This will detect an entangled state when $\left< F \right>_t< 0$. Putting this in the above equation, we get
\begin{eqnarray}
\langle F\rangle_m&<&C_{00}\left(1-\frac{1}{\eta_-}\right)\nonumber \\
&+&\frac{\eta_-}{s(\psi)}\left(\langle H\rangle_m^2+k^2_H-2\langle H\rangle_mk_H\right) \nonumber \\ 
&+&\frac{\eta_-}{s(\psi)}\left(\langle A\rangle_m^2+k^2_A-2\langle A \rangle_mk_A\right)\nonumber \\
&-&\frac{1}{s(\psi)}\left[\langle H\rangle_m^2+\langle A\rangle_m^2\right].  \label{eq32}
\end{eqnarray}
Writing $F$ in terms of the linear witness operator and the nonlinear terms, we get
\begin{eqnarray}
\langle W_{\phi}\rangle_m&<&C_{00}\left(1-\frac{1}{\eta_-}\right) \nonumber \\
&+&\frac{\eta_-}{s(\psi)}\left(\langle H\rangle_m^2+k^2_H-2\langle H\rangle_mk_H\right) \nonumber \\
&+&\frac{\eta_-}{s(\psi)}\left(\langle A\rangle_m^2+k^2_A-2\langle A \rangle_mk_A)\right).\nonumber
\end{eqnarray}
The values of $\left < W_\phi \right>_m$, $\left < H\right>_m$, and $\left< A \right>_m$ which will satisfy the above inequality for a given $\eta_-$ will detect an entangled state, and for that state the loophole would be closed. Although we have derived the condition for loophole closure for a particular case, the method can also be utilized for deriving conditions for other nonlinear entanglement witnesses. 

 Now if we consider the linear witness operator $W_{\phi^+}$ and add the nonlinear term $X=|\phi^+\ket\bra\phi^-|$, the condition for closing the loophole will be
\begin{eqnarray}
\langle W_{\phi^+} \rangle _m &<&\frac{1}{4}\left(1-\frac{1}{\eta_-}\right)+2\eta_-(\langle H\rangle^2_m+\langle A\rangle^2_m).\nonumber   \label{eq10} 
\end{eqnarray}  
Here, $|\phi^-\ket=\frac{1}{\sqrt{2}}\left(|00\ket-|11\ket\right)$. The region above the curved surface in Fig. \ref{gr2} shows the range of $\langle W_{\phi^+} \rangle_m$,  $X_{nl}$, and $\eta_-$, for which the loophole would be closed and an entangled state would be detected. Here, $X_{nl}=\left(\langle H\rangle_m^2+\langle A\rangle_m^2\right)^{\frac{1}{2}}$. The figure shows that the condition for detecting entanglement becomes progressively better as the value of the nonlinear term $X_{nl}^2$ increases. More precisely, for a given value of $\langle W_{\phi^+}\rangle$, an increase in the nonlinear contribution due to $X_{nl}$ allows for the detection of entanglement with lower $\eta_-$.   

\section{NONLINEAR WITNESS OPERATORS FOR BOUND ENTANGLED STATES}
\label{sec-5}

In this section, we begin by identifying nonlinear witness operators for bound entangled states with positive partial transpose. We subsequently show how one can deal with the detection loophole also in this case.

A map $M$, on the space of operators on a Hilbert space, $\mathbb{C}^{d_2}$, which has the property $M(X^\dagger)=M(X)^\dagger$, and which preserves positivity [i.e., if eigenvalues of $X$ are positive, then eigenvalues of $M(X)$ will also be positive], is called a positive map. If we apply $I_{d_1}\otimes M_{d_2}$ on operators on the Hilbert space $\mathbb{C}^{d_1}\otimes\mathbb{C}^{d_2}$ and if the positivity is still preserved, for all $d_1$, then the map is called completely positive. 
\begin{figure*}[t] 
\includegraphics[scale=1.4]{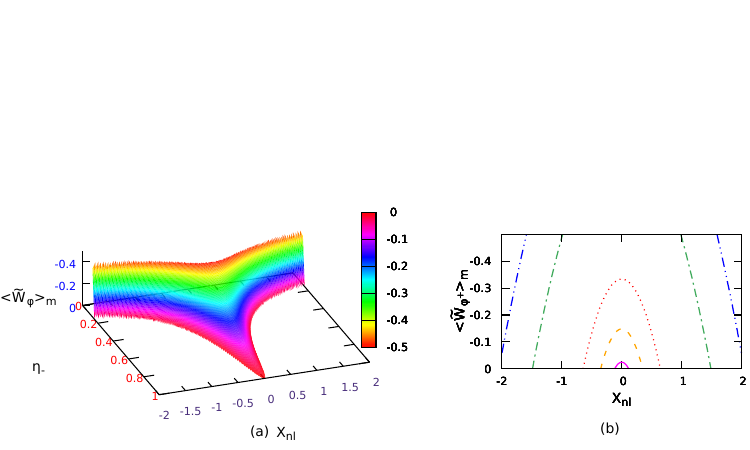}
\caption{Closing the detection loophole for a nonlinear witness to detect entanglement in a bound entangled state. Just like in Fig. \ref{gr2}, the detection of entanglement is determined by the triad, $\langle \widetilde{W}_\phi \rangle_m$, $X_{nl}$, and $\eta_-$. The entanglement detected is of the state $\rho_B$, or of one in a small neighborhood of the same. Despite a possible nonideal detector efficiency, the entanglement of any state in this neighborhood is detected whenever $\langle \widetilde{W}_\phi \rangle_m$, $X_{nl}$, and $\eta_-$ lie in the region above the plotted surface in panel (a), or $\langle \widetilde{W}_\phi \rangle_m$ and $X_{nl}$ lie outside a plotted curve in panel (b) for the corresponding fixed value of $\eta_-$. In panel (a), different values of $\langle \widetilde{W}_\phi \rangle_m$ have been indicated using different colors as shown in the colorbox. The curves in panel (b) have been plotted for some particular values of $\eta_-$. The blue double-dot-dashed, green dot-dashed, red dotted, orange dashed, and magenta continuous lines are, respectively, for $\eta_- = 0.15, 0.2, 0.4, 0.6$, and 0.9. Here again, $X_{nl}=\left( \langle H\rangle_m^2 +\langle A\rangle^2_m \right)^\frac{1}{2}$. All quantities are dimensionless.}
\label{w0}
\end{figure*}
All positive maps behave as completely positive if we restrict their action to separable states on $\mathbb{C}^{d_1}\otimes\mathbb{C}^{d_2}$, and corresponding to every entangled state (say $\rho$) there exists some positive map (say $M_1$) for which $I_{d_1}\otimes M_1(\rho)$ will have at least one negative eigenvalue, for some $d_1$ \cite{Horodecki}. 
Here, $M_1$ is a map on the space of operators on $\mathbb{C}^{d_2}$, and $I_{d_1}$ is the identity map on the space of operators on $\mathbb{C}^{d_1}$.  If an eigenvector corresponding to a negative eigenvalue of $I_{d_1}\otimes M_1(\rho)$ is $|\phi\ket$, then $\widetilde{W}_\phi=\left( I_{d_1}\otimes M_1 \right)^+|{\phi}\ket\bra{\phi}|$ will satisfy the conditions of a witness operator and can detect the state $\rho$.
 Here, $\left(I_{d_1}\otimes M_1 \right)^+$ is defined by the equation $\text{tr}\left[\left(I_{d_1}\otimes M_1\right)^+ \left(O_1\right)O_2\right]=\text{tr}\left[ O_1\left(I_{d_1}\otimes M_1\right)\left( O_2\right)\right]$, for all operators $O_1$ and $O_2$ on $\mathbb{C}^{d_1}\otimes \mathbb{C}^{d_2}$.
 We can now construct a corresponding nonlinear witness operator as 

\begin{eqnarray}
\widetilde{F}=(I\otimes M_1)^+|{\phi}\ket\bra{\phi}|-\frac{1}{s(\psi)g}\left< (I\otimes M_1)^+X\right> \left< \left( (I\otimes M_1)^+X\right)^\dagger\right>,\nonumber\\ \label{eq100} 
\end{eqnarray}
with $g=\max_\sigma\text{tr}[M_1(\sigma)]$, where the maximization is taken over the whole state space, and $X=|\phi\ket \bra \psi |$, where again $|\psi\ket$ is an arbitrary but fixed vector and $s(\psi)$ is the square of the largest Schmidt coefficient of $|\psi\ket$. Let us consider a particular bound entangled state \cite{eita},
\begin{equation}
\rho_B=\frac{2}{7}|{\tilde{\psi}}\ket\bra{\tilde{\psi}}|+\frac{a}{7}\sigma_++\frac{5-a}{7}\sigma_-, \nonumber \end{equation}
where
\begin{eqnarray}
|{\tilde{\psi}}\ket &=&\frac{1}{\sqrt{3}}\left(|{00}\ket+|{11}\ket+|{22}\ket\right), \nonumber \\
\sigma_+&=&\frac{1}{3}\left(|{01}\ket\bra{01}|+|{12}\ket\bra{12}|+|{20}\ket\bra{20}|\right), \nonumber \\
\sigma_-&=&\frac{1}{3}\left(|{10}\ket\bra{10}|+|{21}\ket\bra{21}|+|{02}\ket\bra{02}|\right). \nonumber 
\end{eqnarray}
One can easily check that, for $a \leq4$, $\rho_B$ is PPT.
Now, if we use the map \cite{map},
\begin{eqnarray}
&&M_1
\left(
\left[
\begin{matrix}
a_{11}&a_{12}&a_{13}\\
a_{21}&a_{22}&a_{23}\\
a_{31}&a_{32}&a_{33}
\end{matrix}
\right)
\right]
\nonumber \\ &=&
\left[
\begin{matrix}
a_{11}+a_{33}&-a_{12}&-a_{13}\\
-a_{21}&a_{22}+a_{11}&-a_{23}\\
-a_{31}&-a_{32}&a_{33}+a_{22} 
\end{matrix}
\right],
\label{eq1}
\end{eqnarray}
and find eigenvalues of $I\otimes M_1(\rho_B)$, we can see that it has a negative eigenvalue value for $a>3$. So the state is bound entangled for $3<a\leq 4$.

The eigenvector corresponding to the negative eigenvalue is $|{\phi}\ket=\frac{1}{\sqrt{3}}[|{00}\ket+|{11}\ket+|{22}\ket]$. To construct the nonlinear witness operator as given in \eqref{eq100} let us take $|{\psi}\ket=\frac{1}{2}[|{01}\ket+|{10}\ket+|{12}\ket+|{21}\ket]$. The bound on measured value of $\widetilde{F}$ for detection of entangled states will be similar with the bound given in Eq. \eqref{eq32}, with the only two differences:  $s(\psi)$ will be replaced by $s(\psi)g$, and $H$ and $A$ are now Hermitian - anti-Hermitian decomposition of the operator $(I\otimes M_1)^+X$. Then the corresponding $C_{0H}$ and $C_{0A}$ can be evaluated to be zero and hence $k_{H}$ and $k_{A}$ are also zeros. It can be seen from Eq. \eqref{eq1} that in case of the map $M_1$, $g=2$. Now, decomposition of the linear term, $(I\otimes M_1)^+ |\phi\ket\bra\phi|$, in terms of Gell-Mann matrices, is given by
\begin{equation}
(I \otimes M_1)^+ |\phi\ket\bra\phi |=\sum_{i,j=0}^8 C_{ij}\lambda_i\otimes\lambda_j, \nonumber
\end{equation}
where $\lambda_i$'s are the $3 \times 3$ identity operator [for $i=0$] and Gell-Mann matrices [for $i=1,...,8$]. $C_{ij}$ can be evaluated from the relation
\begin{eqnarray}
&& \text{tr}\left[\left(I\otimes M_1\right)^+ |\phi \ket \bra \phi | (\lambda _i\otimes \lambda _j)\right]=C_{ij}\text{tr}(\lambda _i^2) \text{tr}(\lambda _j^2) \nonumber \\
\Rightarrow \ && \text{tr}\left[|\phi\ket\bra\phi | \left(I\otimes M_1\right) (\lambda _i\otimes \lambda _j)\right]=C_{ij}\text{tr}(\lambda _i^2) \text{tr}(\lambda _j^2) \label{eeq} .
\end{eqnarray}
Now, $M_1$ maps $\lambda_0 \rightarrow 2\lambda_0$. Hence putting $i=j=0$ in \eqref{eeq}, we get $C_{00}=\frac{2}{9}$. Then using inequality \eqref{eq32}, we get
\begin{equation}
\langle \widetilde{F}\rangle_m\leq \frac{2}{9}\left( 1-\frac{1}{\eta_-}\right)+2(\eta_--1)\left[\langle H \rangle_m^2+\langle A \rangle_m^2\right]. \nonumber
\end{equation}
Representing the above relation in terms of the linear witness operator, we get
\begin{equation}
\langle \widetilde{W}_\phi\rangle_m\leq \frac{2}{9} \left(1-\frac{1}{\eta_-}\right)+2\eta_-\left[\langle H\rangle_m^2+\langle A\rangle_m^2 \right]. \label{eq101}
\end{equation}
This is the condition for detecting bound entangled states in real experiments in which the detector does not work ideally. The boundary beyond which the nonlinear witness operator can detect entanglement of $\rho_B$ is shown in Fig. \ref{w0}. We can see that as the measured value of $X_{nl}^2$ increases from 0 to 4, i.e., the value of $X_{nl}$ increases from 0 to 2 or decreases from 0 to -2, the chance of detecting entanglement increases, i.e., it increases with increase in the measured value of nonlinear terms. 
For example, suppose that the value of \(\langle \widetilde{W}_\phi \rangle_m\) is zero.
Then, if the nonlinear term \(X_{nl}\) is 0.71, detection of entanglement is possible for 
\(\eta_- \gtrsim 0.37\). For the same value of \(\langle \widetilde{W}_\phi \rangle_m\), if the nonlinear term 
\(X_{nl}\) attains a higher value of, say, 1.0, the same detection is possible for the larger range of the efficiency, viz., \(\eta_- \gtrsim 0.28\).

We can also conclude by observing the figures or from relation \eqref{eq101} that the nonlinear witness constructed for detecting the bound entangled state is better than its corresponding linear witness for any nonzero value of the lost event efficiency.

\section{CONCLUSION}
\label{sec6}

We found conditions for detection of entanglement in bipartite quantum
states using nonlinear witness operators in situations where the detectors
have nonideal, but known, efficiency. The method is related to the way
that the detection loophole is dealt with in experiments looking for
violation of Bell inequalities, and for detection of entanglement using
linear entanglement witnesses. While the method followed can be
generalized to several other situations, we have first dealt with the case
of detecting entangled states with a nonpositive partial transpose by
using a nonlinear witness operator related to the positive partial
transpose criterion. We have then found nonlinear entanglement witnesses
for a bound entangled state, and have subsequently derived conditions for
it to perform the detection in presence of errors. In both the cases, the
nonlinear witnesses turn out to be more efficient in detecting
entanglement, even for nonideal efficiencies, than their linear
counterparts.

\acknowledgments

We acknowledge useful discussions with Aditi Sen(De).


\begin{thebibliography}{9}
\bibitem{HHHH}
R. Horodecki, P. Horodecki, M. Horodecki, and K. Horodecki,
Rev. Mod. Phys. \textbf{81}, 865 (2009).
\bibitem{sree}
 S. Das, T. Chanda, M. Lewenstein, A. Sanpera, A. Sen De, and U. Sen, arXiv:1701.02187.
\bibitem{tele}
C. H. Bennett, G. Brassard, C. Cr\'epeau, R. Jozsa, A. Peres, and W. K. Wootters, Phys. Rev. Lett. \textbf{70}, 1895 (1993).
\bibitem{dc}
C. H. Bennett and S. J. Wiesner, Phys. Rev. Lett. \textbf{69}, 2881 (1992).
\bibitem{ekert}
A. K. Ekert, Phys. Rev. Lett. \textbf{67}, 661 (1991).
\bibitem{Peres}
A. Peres, Phys. Rev. Lett. \textbf{77}, 1413 (1996).
\bibitem{Horodecki}
M. Horodecki, P. Horodecki, and R. Horodecki, Phys. Lett. A. \textbf{223}, 1 (1996).
\bibitem{entr1}
R. Horodecki and P. Horodecki, Phys. Lett. A. \textbf{194}, 147 (1994).
\bibitem{entr2}
M. A. Nielsen and J. Kempe, Phys. Rev. Lett. \textbf{86}, 5184 (2001).
\bibitem{Bell}
J.S. Bell, \textit{Speakable and unspeakable in
quantum mechanics} (Cambridge University Press., New York, 1989).
\bibitem{witness}
B. M. Terhal, Physics Letters A \textbf{271}, 319 (2000); D. Bru{\ss}, J. I. Cirac, P. Horodecki, F. Hulpke, B. Kraus, M. Lewenstein, and A. Sanpera, J. Mod. Opt. \textbf{49}, 1399 (2002); O. G\"uhne and G. T\'oth, Physics Reports \textbf{474}, 1 (2009).
\bibitem{exp}
B. P. Lanyon, M. Zwerger, P. Jurcevic, C. Hempel, W. D\"ur, H. J. Briegel, R. Blatt, and C. F. Roos,
Phys. Rev. Lett. \textbf{112}, 100403 (2014);
C. J. Ballance et al., Nature  \textbf{528},  384 (2015);
M. Cramer, A. Bernard, N. Fabbri, L. Fallani, C. Fort, S. Rosi,
F. Caruso, M. Inguscio, and M. B. Plenio, Nat. Commun. \textbf{4},
2161 (2013);
The BIG Bell Test Collaboration, Nature \textbf{557}, 212 (2018).
\bibitem{wit_exp}
N. Friis et al., Phys. Rev. X \textbf{8}, 021012 (2018); 
A. Stute, B. Casabone, P. Schindler, T. Monz, P. O. Schmidt, B. Brandst\"atter, T. E. Northup and R. Blatt, Nature \textbf{485}, 482 (2012);
H.-S. Zhong et al., Phys. Rev. Lett. \textbf{121}, 250505 (2018);
X.-L. Wang et al., Phys. Rev. Lett. \textbf{117}, 210502 (2016);
L. DiCarlo, M. D. Reed, L. Sun, B. R. Johnson, J. M. Chow, J. M. Gambetta, L. Frunzio, S. M. Girvin, M. H. Devoret and R. J. Schoelkopf, Nature \textbf{467}, 574 (2010);
J. M. Chow et al., Nat. Commun. \textbf{5}, 4015 (2014).
\bibitem{hahn}
G. F. Simmons, \textit{Introduction to Topology and Modern Analysis} (Krieger Publishing Company, Malabar, 2003).
\bibitem{CHSH}
J. F. Clauser, M. A. Horne, A. Shimony, and R. A. Holt,
Phys. Rev. Lett. \textbf{23}, 880 (1969).
\bibitem{X}
P. Pearle, Phys. Rev. D \textbf{2}, 1418 (1970);
J. F. Clauser, and M. A. Horne, Phys. Rev. D \textbf{10}, 526 (1974);
E. Santos, Phys. Rev. A \textbf{46}, 3646 (1992).
\bibitem{two}
N. Gisin and B. Gisin, Phys. Lett. A \textbf{260}, 323 (1999);
C. Branciard, Phys. Rev. A \textbf{83}, 032123 (2011);
Y. Lim, M. Paternostro, M. Kang, J. Lee, and H. Jeong, Phys. Rev. A \textbf{85}, 062112 (2012); 
K. F. P\'al and T. V\'ertesi, Phys. Rev. A \textbf{92}, 022103 (2015);
 J. Szangolies, H. Kampermann, and D. Bru{\ss}, Phys. Rev. Lett. \textbf{118}, 260401 (2017);
 E. Z. Cruzeiro and N. Gisin, Phys. Rev. A \textbf{99}, 022104 (2019).  
\bibitem{exp_bell}
L. K. Shalm et al., Phys. Rev. Lett. \textbf{115}, 250402 (2015);
M. Giustina et al., Phys. Rev. Lett. \textbf{115}, 250401 (2015);
D. Rauch et al.,Phys. Rev. Lett. \textbf{121}, 080403 (2018);
W. N. Plick, F. Arzani, N. Treps, E. Diamanti, and D. Markham; Phys. Rev. A \textbf{98}, 062101 (2018);
M. Ansmann, H. Wang, R. C. Bialczak, M. Hofheinz, E. Lucero, M. Neeley, A. D. O’Connell, D. Sank, M. Weides, J. Wenner, A. N. Cleland, and J. M. Martinis, Nature \textbf{461}, 504 (2009);
B. Hensen et al., Nature \textbf{526}, 682 (2015); 
J. Hofmann, M. Krug, N. Ortegel, L. G\'erard, M. Weber, W. Rosenfeld, and H. Weinfurter, Science \textbf{337}, 72 (2012).
\bibitem{Bruss}
P. Skwara, H. Kampermann, M. Kleinmann, and D. Bru$\beta$, Phys. Rev. A \textbf{76}, 012312(2007). 
\bibitem{nonlinear1}
O. G\"uhne and N. L\"utkenhaus, Phys. Rev. Lett. \textbf{96}, 170502 (2006).
\bibitem{nonlinear2}
O. G\"uhne, M. Mechler, G. T\'oth, and P. Adam, Phys. Rev. A \textbf{74}, 010301(R) (2006);
O. G\"uhne and N. L\"utkenhaus, J. Phys.: Conf. Ser. \textbf{67}, 012004 (2007);
C.-J. Zhang, Y.-S. Zhang, S. Zhang, and G.-C. Guo, Phys. Rev. A \textbf{76}, 012334 (2007); 
T. Moroder, O. G\"uhne, and N. L\"utkenhaus, Phys. Rev. A \textbf{78}, 032326 (2008);
M. A. Jafarizadeh, M. Mahdian, A. Heshmati, and K. Aghayar, Eur. Phys. J. D \textbf{50}, 107 (2008);
M. A. Jafarizadeh, A. Heshmati, and K. Aghayara, arXiv:0910.5413;
M. Kotowski, M. Kotowski, and M. Ku\'s, Phys. Rev. A \textbf{81}, 062318 (2010);
J.-Y. Wu, H. Kampermann, D. Bru{\ss}, C. Kl\"ockl, and M. Huber, Phys. Rev. A \textbf{86}, 022319 (2012);
M. Agnew, J. Z. Salvail, J. Leach, and R. W. Boyd, arXiv:1210.1054;
J. M. Arrazola, O. Gittsovich, J. M. Donohue, J. Lavoie, K. J. Resch, and N. L\"utkenhaus, Phys. Rev. A \textbf{87}, 062331 (2013);
M. Oszmaniec and M. Ku\'s, Phys. Rev. A \textbf{90}, 010302 (2014);
K. Aghayar, A. Heshmati, and M. A. Jafarizadeh, arXiv:1507.06979;
K. Lemr, K. Bartkiewicz, and A. \v Cemoch, Phys. Rev. A \textbf{94}, 052334 (2016);
X.-y. Chen and L.-z. Jiang, arXiv:1901.02312.
\bibitem{nonlinear3}
J. Maziero and R. M. Serra, Int. J. Quant. Inf. \textbf{10}, 1250028 (2012);
M. Li, T.-J. Yan and S.-M. Fei, J. Phys. A: Math. Theor. \textbf{45}, 035301 (2012);
Z.-H. Ma, Z.-H. Chen, and J.-L. Chen, arXiv:1104.0299;
J. Bowles, M. T. Quintino, and N. Brunner, Phys. Rev. Lett. \textbf{112}, 140407 (2014). 
 \bibitem{bound}
P. Horodecki, Phys. Lett. A \textbf{232}, 333 (1997); M. Horodecki, P. Horodecki, and R. Horodecki, Phys. Rev. Lett. \textbf{80}, 5239 (1998).
\bibitem{eita}
P. Horodecki, M. Horodecki, R. Horodecki, Phys. Rev. Lett. \textbf{82}, 1056 (1999).
\bibitem{1989}
R. F. Werner, Phys. Rev. A \textbf{40}, 4277 (1989).
\bibitem{Ppt}
A. Sanpera, R. Tarrach and G. Vidal, Phys. Rev. A \textbf{58}, 826 (1998).
\bibitem{map}
M. D. Choi, Linear Algebra Appl. \textbf{12}, 95 (1975).
\end{thebibliography}
\end{document}